# An atomic force microscope integrated with a helium ion microscope for correlative nanocharacterization

Santiago H. Andany[1], Gregor Hlawacek[2], Stefan Hummel[3], Charlène Brillard[1], Mustafa Kangül[1] and Georg E. Fantner*[1]

Address: [1]Laboratory for Bio- and Nano-Instrumentation, Swiss Federal Institute of Technology Lausanne (EPFL), Lausanne CH-1015, Switzerland, [2]Institute of Ion Beam Physics and Materials Research, Helmholtz-Zentrum Dresden-Rossendorf, Dresden 01328, Germany and [3]GETec Microscopy GmbH, Vienna 1220, Austria

Email: Georg E. Fantner* – georg.fantner@epfl.ch

* Corresponding author

## Abstract

In this work, we report the integration of an atomic force microscope (AFM) into a helium ion microscope (HIM). The HIM is a powerful instrument, capable of sub-nanometer resolution imaging and machining of nanoscale structures, while the AFM is a well-established versatile tool for multiparametric nanoscale characterization. Combining the two techniques opens the way for unprecedented, in situ, correlative analysis at the nanoscale. Nanomachining and analysis can be performed without contamination of the sample and environmental changes between processing steps. The practicality of the resulting tool lies in the complementarity of the two techniques. The AFM offers not only true 3D topography maps, something the HIM can only provide in an indirect way, but also allows for nanomechanical property mapping, as well as for electrical and magnetic characterization of the sample after focused ion beam materials modification with the HIM. The experimental setup is described and evaluated through a series of correlative experiments, demonstrating the feasibility of the integration.

## Introduction

Shortly after the invention of the atomic force microscope (AFM) in 1986 [1], efforts were made towards combining this scanning probe microscopy technique with electron beam and ion beam techniques for correlative nano-characterization and nano-fabrication. The motivation was driven by the new opportunity to investigate and transform features in situ, with complementary techniques, thus revealing maximum information without breaking the vacuum. The scanning electron microscope (SEM) was first combined with scanning tunneling microscopy (STM) [2,3], allowing visual observation at the tip-sample interaction point with the SEM. Later, Ermakov et al. [4] successfully integrated an AFM into an SEM for the first time, enabling correlative imaging on electrically insulating samples. In this first attempt, the readout of cantilever deflection was achieved using the electron beam itself. Shortly after, better performing combined setups were described utilising more conventional self-sensing [5] and optical [6] techniques for the readout of cantilever deflection. Since then, more advanced and versatile combined instruments have been proposed for a broad spectrum of applications in nano-characterization and nano-fabrication inside SEM and focused ion beam (FIB) setups [7–11].

Given the extent of the interest sparked by SEM/FIB-AFM systems, it is reasonable to assume that the most recent ion beam microscope, the helium ion microscope (HIM), would present as a serious contender for use in combined setups, in conjunction with AFM. Introduced by Ward et al. [12], the HIM's imaging capability surpasses that of the SEM in terms of lateral resolution, depth of field, surface sensitivity, and ability to image electrically insulating samples [13]. Furthermore, nano-structuration with noble gas ions can yield sub 10 nm structures without unwanted metal ion implantation, a sizeable advantage over traditional gallium ion FIBs. The resulting combined AFM-HIM instrument would, therefore, profit from the sub-nanometer lateral resolution of the HIM and atomic resolution in the vertical axis with the AFM, proving particularly powerful for high-resolution correlative characterization of non-conductive samples.

With the integrated electron flood gun (FG) of the HIM providing charge neutralization, uncoated polymers and biological samples can be imaged with high resolution while the AFM would bring complementary information such as laterally resolved mechanical properties. These multiparametric measurements have previously been difficult to obtain as sample preparation of such samples for SEM or TEM are often incompatible with the needs of high-resolution AFM measurements.

AFM is also useful in assisting helium ion beam lithography. Many resists, including Poly(methyl methacrylate) (PMMA), have higher sensitivities to helium ion irradiation than to electron irradiation in terms of charge per area [14]. Patterning resolution down to 4 nm has been demonstrated on HSQ resist [15], surpassing electron beam lithography, which greatly suffers from the proximity effect. In a combined AFM-HIM setup, the AFM could be used, in situ, in between exposures to assess the shrinkage, stiffness change or sputtering of the resist. More applications such as conductive AFM, piezo-force microscopy or magnetic force microscopy are within reach of the presented technology and would make AFM-HIM appealing to the microelectronics and materials research community.

## Results

**Instrumentation**

Spatial constraints inside SEMs and ion microscopes often dictate the feasibility of the integration of the AFM. Compact AFM setups have to fit around the host microscope as not to hinder excessively its capabilities. The reported AFM integration is depicted in Figure 1. The prototype AFM scan head is designed explicitly for correlative analysis inside electron and ion-beam microscopes. It consists of a compact flexure-based assembly made of grade 5 titanium (Ti 6Al-4V) with 3-axis of motion actuated by stack-piezo actuators, offering an achievable scan range of 30x30x12 µm. The instrument uses silicon piezo-resistive self-sensing cantilever probes with single crystal diamond tips (SCL-SensorTech Fabrication GMBH, Vienna, Austria), eliminating the need for a voluminous optical readout. To maneuver the AFM relative to the sample and the ion beam, the AFM is mounted onto a coarse stage consisting of a homebuilt XY stick-slip positioner which in turn is attached to a vertical approach mechanism built around a linear, stick-slip piezo actuator (Picomotor$^{TM}$ 8301-UHV, Newport Corporation, CA, USA). The AFM assembly tilts together with the sample stage. The three orthogonal translational degrees of freedom of the sample are decoupled from the AFM coarse positioning stage, as shown in Figure 1a. The integration of the AFM into the HIM requires no alteration of the HIM microscope stage. The AFM assembly is positioned onto the HIM cradle and secured with set-screws pressing firmly on the sides of the cradle. Electrical connections necessary for AFM operation are cabled through a CF40 flange. After opening the microscope door, the AFM head can be removed seamlessly from the chamber for cantilever exchange thanks to a spring-loaded kinematic mount.

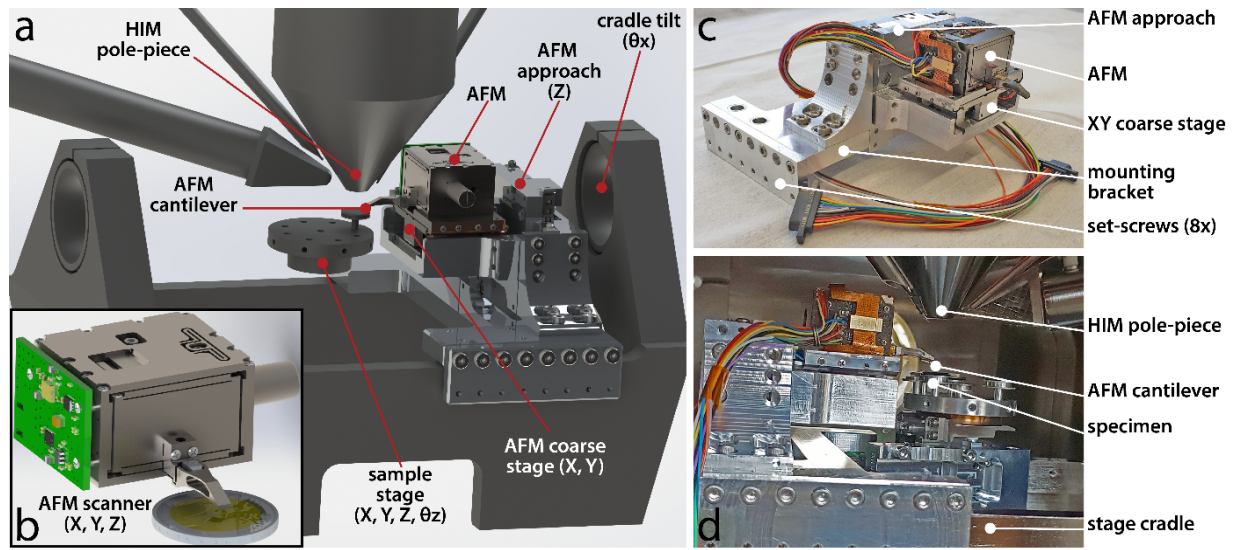

**Figure 1:** AFM assembly and integration inside a Zeiss ORION NanoFab Helium Ion Microscope. a) Simplified CAD rendering of the AFM assembly mounted onto the HIM cradle and b) detailed view of the AFM scan head and a 2€ coin for scale. c) Annotated photograph of the AFM assembly and d) after being mounted inside the chamber of the HIM.

The AFM and motorized coarse stages are controlled with a home-made AFM software [16], a standalone FPGA (USB-7856R OEM, National Instruments, Austin TX, USA), a high-voltage piezo amplifier (Techproject, Vienna, Austria) and a stick-slip controller (8742-4 PicomotorTM drive, Newport Corporation).

## Experimental results

The system has been experimentally tested on a variety of sample surfaces in contact and off-resonance imaging modes, demonstrating the feasibility of the integration through a series of three experiments. Correlative AFM and HIM imaging is demonstrated in Figure 2 by imaging silicon nano-pillars [17]. The HIM offers a large field of view, which allows for the cantilever to be navigated onto the region of interest (Figure 2b and 2c) to perform AFM topography imaging (Figure 2d).

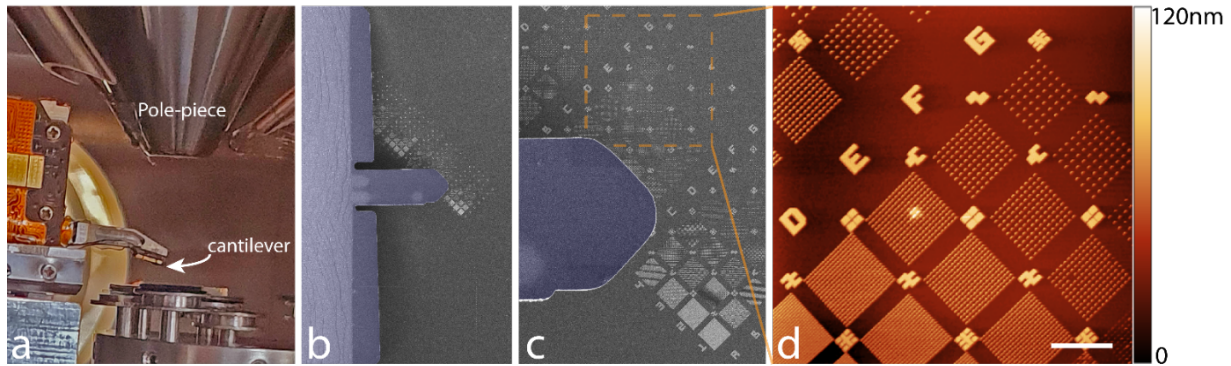

**Figure 2:** Correlative imaging in process on silicon micro-pillars. a) Optical image showing how the AFM cantilever is positioned at the end of a low-profile, overhanging structure that fits between the pole piece and the sample. b) & c) The cantilever (colourized in purple on the HIM images) can be navigated by making use of the large FOV image provided by the HIM. d) AFM height image of Si nano-pillars taken in off-resonance tapping mode. Scale bar 5 µm.

PMMA has traditionally been used as a positive resist with electron beam lithography. Helium ion beam lithography has emerged as a powerful technique to achieve even smaller feature size thanks to higher resist sensitivity, reduced proximity effect and small spot size [15]. Upon ion beam exposure, chain scission occurs leaving the exposed region soluble in a suitable developer. Very high ion doses also break short side chains that later cross-link, allowing PMMA to be also used a negative resist [18]. Chain scission leads to volume loss through the release of gas molecules, and this leads to the shrinkage of exposed PMMA [19], which can be easily quantified using AFM. In a second experiment, we tested the effect on a PMMA thin film as it is exposed to different doses under the focused He ion beam. Figure 3a and Figure 3b show two AFM topography images of PMMA that has been exposed to a dose of $1\times10^{13}$ cm$^{-2}$ and $3\times10^{13}$ cm$^{-2}$ 30 keV He ions respectively, and the corresponding height profiles of the irradiated PMMA surface.

Focused ion beam damage and implantation can hinder the imaging and nanofabrication capabilities of the HIM [20] and studying these local defects created at the micro and nanoscale can provide valuable information towards understanding these limitations. For example, a focused helium ion beam can locally destroy the crystalline structure of silicon and lead to the growth of amorphous silicon bubbles at the surface [21]. Furthermore, focused helium ion beam exposure inside a HIM can be used as a way of locally replicating the strong irradiation conditions found in nuclear fission and fusion reactors, to study the response of structural materials used in the reactors [22]. We characterized the defects caused by He ion exposure in a correlative AFM-HIM. Amorphous silicon bubbles are created on a crystalline silicon substrate through point exposition with the HIM at 25 kV and 14 pA using doses between $4.2\times10^8$ and $4.2\times10^9$ He ions (see Figure 3). He ions penetrate deep into the silicon and lead to the formation of micro and nano bubbles that coalesce and ultimately result in the formation of a large silicon bubble in the amorphized silicon. The resulting 3x3 bubble grid is imaged with HIM (Figure 3c) and AFM (Figure 3d) to reveal the height and volume of the features.

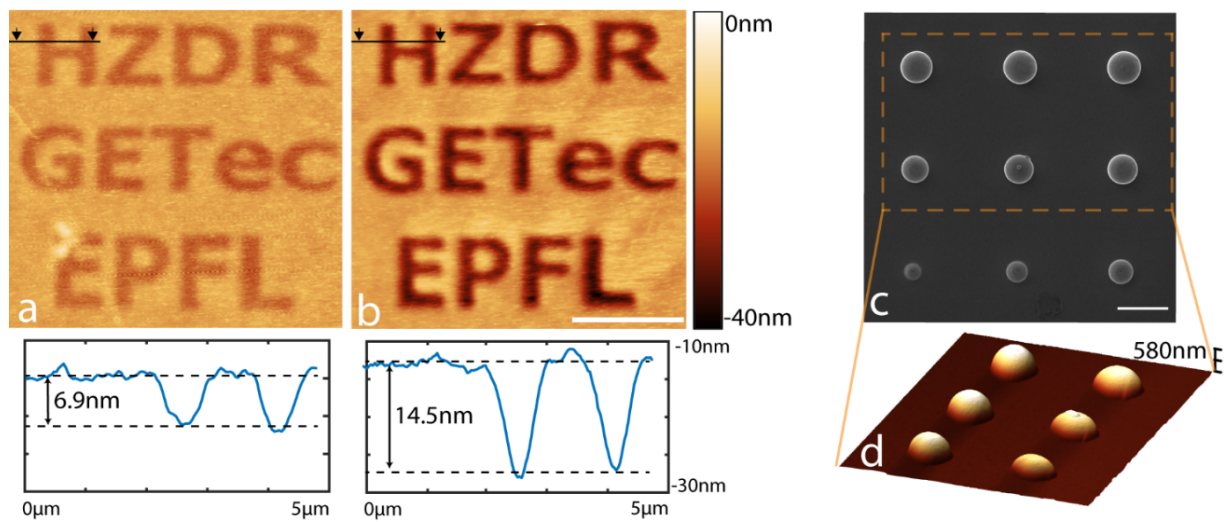

**Figure 3:** AFM height images of Poly(methyl methacrylate) after exposure to a) 1x10$^{13}$ and b) 3x10$^{13}$ He ion cm$^{-2}$. Image is taken in off-resonance tapping mode, scale bar 4 μm. c) Silicon bubbles imaged with HIM and d) AFM (off-resonance tapping mode, scale bar 1 μm.)

## Discussion

For the successful integration of two different microscopy techniques, they should be both complementary and compatible. Techniques should, on the one hand, be sufficiently different so that the combination creates real added value, but on the other hand, the application space of the techniques should have sufficient overlap so that a meaningful correlation can be established in space and time. One requirement for compatibility is that the AFM can operate in the ultra-high vacuum (UHV) environment, a prerequisite for the HIM. This requirement puts additional restrictions on the AFM. In our AFM design, we accounted for this already in the mechanical design (avoidance of trapped air pockets, lubrication-free, UHV compatible motors) as well as in the assembly by using wherever possible Kapton flex-PCBs or low outgassing Teflon coated wires. We should note, however, that our AFM system is not compatible with baking the system at high temperatures above 100°C since this would result in irreversible damage to the piezo actuators.

Another requirement for compatibility is that the two techniques can use samples prepared in the same way. For AFM and HIM, this is particularly advantageous since both can image non-conductive samples at very high resolution without charging. This is essential for correlative mechanical property and HIM surface imaging, and it is a clear advantage of AFM-HIM compared to AFM-SEM, where a conductive coating is often necessary for high-resolution SEM imaging.

The other aspect necessary for a useful integration of two techniques is that they are sufficiently complementary to each other to warrant the additional effort. While both AFM and HIM can yield very high-resolution images, the two techniques do have very different strengths. The HIM, for example, has a very good lateral resolution and a large depth of field, which makes it well suited for imaging high aspect ratio structures. The Z-resolution of the method, however, is less accurate, since the height of objects has to be back-calculated from two tilted images. AFM, on the other hand, has its highest resolution in the Z-direction, and profiles or volumes can be accurately extracted (see Figure 3d). The depth of field is, however, limited and the maximum slope of the sample that can be faithfully measured is dictated by the aspect ratio of the tip [23]. The true strength of the integrated setup is the combination of sample modification by the He ion beam and the multiparametric characterization of sample properties using AFM. In Figure 3, we showed a basic application where we characterized the effect of ion-beam radiation on the topography of the photoresist PMMA. Many more examples can be envisioned. The He ion beam is known to change the mechanical [24], electrical [25], and magnetic properties of materials [26]. AFM can be used to measure mechanical properties using contact resonance [27,28] or off-resonance tapping techniques [29] with very high resolution. Magnetic properties of nanostructures can be measured using magnetic force microscopy (MFM) [30], and a host of AFM techniques are available to measure electrical properties of samples (conductive AFM (cAFM) [31], scanning capacitance microscopy (SCM) [32], spreading resistance microscopy (SSRM) [33] etc.). While the implementation of these different imaging modes will require some additional modifications to our existing instrument, the path towards achieving such a truly multi-physics characterization and manipulation tool by combining advanced AFM with HIM can clearly be envisioned.

One aspect where HIM and AFM are, however, not well matched is in the image acquisition time. The relatively long time required for an AFM image (several minutes) has been a severe disadvantage when combining it with other electron or ion-beam microscopes. The same limitation exists for the combination of AFM and HIM. While much progress has been made towards increasing the imaging speed of AFM [34–38], most of this progress has been limited to imaging in liquid, due to the inherent bandwidth limitation of cantilevers when using them in dynamic mode in vacuum. Recent signs of progress in cantilever materials have shown the potential to increase the imaging speed of AFM also in ambient air or vacuum [39–41]. These approaches could also be implemented for the combined AFM-HIM instrument, thereby holding promise for interactive use of AFM and HIM at similar size and time-scales.

## Conclusion

We have demonstrated the integration of an atomic force microscope into a helium ion microscope. Correlative measurements of AFM topography with He ion imaging and modification demonstrate the feasibility of this integration. The complementarity of the two methods in terms of vertical and lateral resolution, nanoscale machining, and measurement of physical properties of the sample will allow for a multi-physics investigation in many areas of materials science and technology, such as energy materials, magnetic nanostructures, and (bio-) composites.

# Experimental

All AFM measurements were taken using silicon piezo-resistive self-sensing cantilevers (PRS-L100-F500-SCD-PCB, SCL-SensorTech Fabrication GMBH, Vienna, Austria) with single crystal diamond tips (radius around 15 nm), a spring constant around 100 N/m, and a footprint of 110x48 µm. Imaging gains on the homemade controller were adjusted as high as possible before significant oscillations were seen. AFM images were processed in the software Gwyddion [42]. Pixels were squared to account for X-Y pixel size mismatch when necessary, the background was flattened and a conservative de-noising filter was applied. Finally, hysteresis correction was performed in MATLAB using closed-loop sensor data obtained prior to imaging on the AFM scan head.

For Figure 2, the AFM image shown was performed at 300 mHz line rate at a resolution of 1024 pixels and 512 lines and at a scan range of 30x30 µm. The imaging mode used was off-resonance tapping (ORT) at a tapping rate of 2 kHz and a tapping amplitude of 600 nm. For Figure 2a and Figure 2b, the images were taken in contact mode at 500 mHz line-rate and 1 Hz line rate respectively. Additionally to the processing detailed above, the 2 images were cropped and rotated to obtain the final images (original images are 17.8x17.8 µm and 16.6x16.6 µm respectively and each are 512x512 pixels). An additional 2-dimensional FFT filtering was applied to correct for the main mechanical vibrations in the 2 original images. The AFM image in Figure 3d is obtained in ORT at 2 kHz tapping rate, 600 nm amplitude and 200 mHz linerate. The scan range is 9.7x7.3 µm and the image size is 512x386 pixels.

## Acknowledgements

The authors acknowledge the support of Jeff Markakis for the mechanical design of the AFM scanner and Christian Schwalb for conceptual discussions.

## Funding

This work was funded by the Swiss National Science Foundation (the Swiss National Science Foundation through grant 200021_182562), the European Union Eurostars Program (Eurostars E!8213), the H2020 IONS4SET project (H2020 Grant number: 688072) and the Ion Beam Center of the Helmholtz Zentrum Dresden Rossendorf through GATE proposal 19001761.